\documentclass[12pt]{article}
\usepackage[utf8]{inputenc}

\usepackage{amsmath}
\usepackage{biblatex}
\usepackage{graphicx}
\usepackage[vlined,ruled,linesnumbered]{algorithm2e}
\usepackage{geometry}
\usepackage[hidelinks]{hyperref}
\usepackage{enumitem}
\usepackage{changepage}
\usepackage{caption}
\title{Sorting Out New York City's Trash Problem}
\author{Steven DiSilvio, Anthony Ozerov, Leon Zhou}
\date{January 2022}

\begin{document}

\graphicspath{{.}}  % Place your graphic files in the same directory as your main document
\DeclareGraphicsExtensions{.pdf, .jpg, .tif, .png}

%%%%%%%%%%% Begin Summary %%%%%%%%%%%

\maketitle

\begin{abstract}

\noindent 

%To achieve zero waste and improve public health and sanitation in New York City, innovative policies tailored to the city's unique urban landscape are necessary. Our proposed programs include the Dumpster and Compost Accessibility Program, which places affordable dumpsters near fire hydrants and provides compost bins to households to increase composting rates, and the Pay-As-You-Throw Program, which requires single/two-family households to purchase stickers for refuse bags and incentivizes waste sorting. Using weighted multi-objective optimization, we determine the optimal sticker price that balances the city's priorities. These programs will increase diversion rates, reduce littering, and potentially save the city money.

\noindent To reduce waste and improve public health and sanitation in New York City, innovative policies tailored to the city's unique urban landscape are necessary.
The first program we propose is the Dumpster and Compost Accessibility Program. This program is affordable and utilizes dumpsters placed near fire hydrants to keep waste off the street without eliminating parking spaces. It also includes legal changes and the provision of compost bins to single/two-family households, which together will increase composting rates.
The second program is the Pay-As-You-Throw Program. This requires New Yorkers living in single/two-family households to purchase stickers for each refuse bag they have collected by the city, incentivizing them to sort out compostable waste and recyclables. We conduct a weighted multi-objective optimization to determine the optimal sticker price based on the City's priorities. Roughly in proportion to the price, this program will increase diversion rates and decrease the net costs to New York City's Department of Sanitation.
In conjunction, these two programs will improve NYC's diversion rates, eliminate garbage bags from the streets, and potentially save New York City money.
%We propose two complementary programs to solve these problems.

%To bring New York City to zero waste sent to landfills and to eliminate public health problems and littering, creative policies must be designed and implemented. This must be done while considering the varied nature of New York's demographics and urban landscape. We propose two complementary programs to solve these problems.

%The first program is the Dumpster and Compost Accessibility Program. This program replaces the unsightly bags which are the necessary result of current collection practices with dumpsters that keep waste out of sight, out of mind, and out of rodents' mouths. It also includes legal changes and the provision of compost bins to single/two-family households, which together will increase composting rates to those of recycling. This program is easily affordable.

\end{abstract}

\tableofcontents

\newpage
\section{Introduction}

\begin{figure}[!h]
\centering
\includegraphics[width=0.5\textwidth]{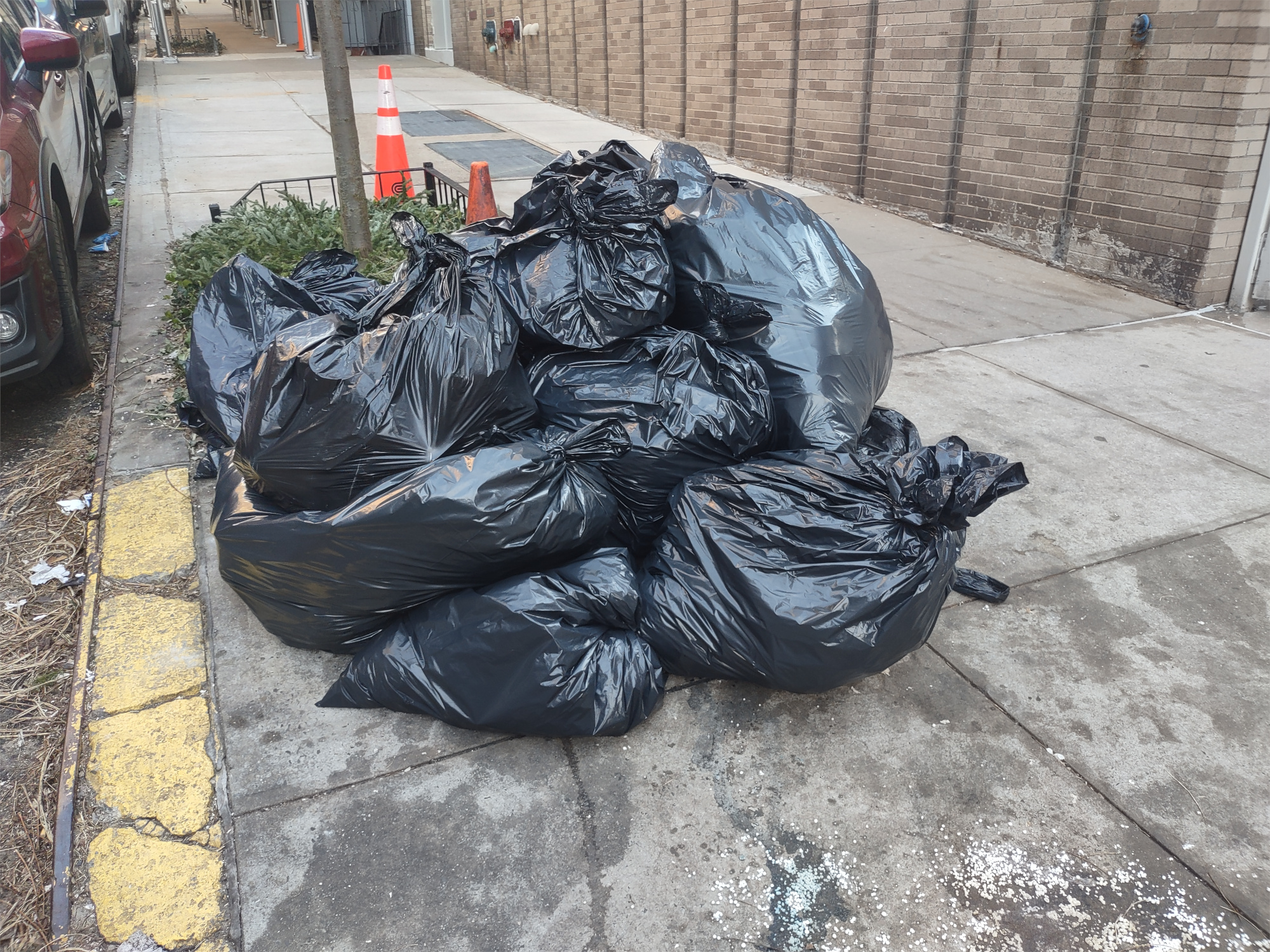}
\caption{Bags of trash left outside for curbside pickup near the present authors. Rats have been observed in this area.} \label{fig:bags}
\end{figure}

\noindent Among New York City (NYC) residents, trash is a common source of complaints about the city \cite{nytimes_trash}. Trash is a very visible problem, and many images similar to Figure \ref{fig:bags} are a daily sight on NYC streets. Furthermore, in 2017 only about 1.2\% of the 1 million tons of organic waste produced by the city was properly composted \cite{goldenberg_muoio_2020}. Despite this, NYC also spends about $\$500$ per household for waste management, which is significantly more than other cities \cite{500household}. Clearly, this is a problem that must be solved by the City. Existing research on municipal solid waste management has focused on:
\begin{itemize}
    \item Modeling different final destinations for waste, their environmental externalities, and the monetary value of these externalities\cite{externalities}.
    \item Improving upon the routing of collection trucks. This was notably studied in NYC in 2018 \cite{wastezones}, and significant improvements have been made.
    \item Using multicriteria decision analysis to help choose a solid waste management plan \cite{mcda}.
    \item The effects of incentive schemes, such as Pay-As-You-Throw (PAYT), on recycling levels and costs \cite{elasticity}.
\end{itemize}

\noindent We propose several different solid waste management policies for NYC, and organize them into two programs for analysis. We examine their effects (environmental, economic, aesthetic) and costs. As there is limited data on the current destinations of NYC's waste (landfills, incinerators, recycling, anaerobic digestion, etc.), we do not believe it is feasible to conduct a full analysis of where NYC can or should send its waste. Due to recent advances in dumpster truck routing in NYC \cite{wastezones}, as well as our relative lack of data compared to NYC Department of Sanitation (DSNY) route planners, we also do not believe it is feasible to significantly improve upon vehicle routing. Instead, we propose simple interventions, including a program to clean up NYC's streets and improve composting rates, and a PAYT policy to increase recycling and composting rates.

NYC is a tale of two cities in terms of how residents interact with their refuse and recycling. Residents in apartment complexes place their waste in common bins, often on every floor of a building \cite{law}. In the case of one of the present authors, this waste is then dealt with by the building management, which places it in bags on the street. A PAYT policy impossible in such a setting. However, in single/two-family homes, common on Staten Island and in Queens, residents deal much more directly with their waste, as they place it in bins or bags on the street. Our policies take these differences into account at the zoning district level instead of the community district or borough level, as many community districts (e.g. Queens' community district 12) have a mix of apartment buildings and single/two-family homes. We study our PAYT policy in the context of Queens, as it has a mix.

We define ``waste'' and ``trash'' as everything that is disposed of by residents, and ``refuse'' as anything that is not compostable or recyclable.

\section{Models and Policies}
\subsection{Dumpster and Compost Accessibility Program}

\paragraph{Problems} One problem in NYC is the lack of designated street containers for residential waste. NYC has few alleyways (where dumpsters are usually placed). This leads to residential waste being placed in bags on the street \cite{beyer_2020}, as in Figure \ref{fig:bags}. There is little data on the relationship between trash bags and rat populations, but the bags of trash left on the street are still generally considered unsightly.

This leads to data like the fact that, in Manhattan in 2017, roughly 0\% of compostable waste was properly disposed of while 50\% of recycling was properly disposed of \cite{tonnage,subsort}. This is a major problem, as in NYC compostable waste is responsible for over a third of all waste \cite{tonnage}. For commercial districts, there are already laws in NYC requiring them to separate compost and hire private transport to dispose of it \cite{compostlaw}, so the main problem resides in residential units. Compost bins (unlike recycling bins) are few and far between, and many buildings have no composting service.

\begin{figure}[!ht]
\centering
\includegraphics[width=0.6\textwidth]{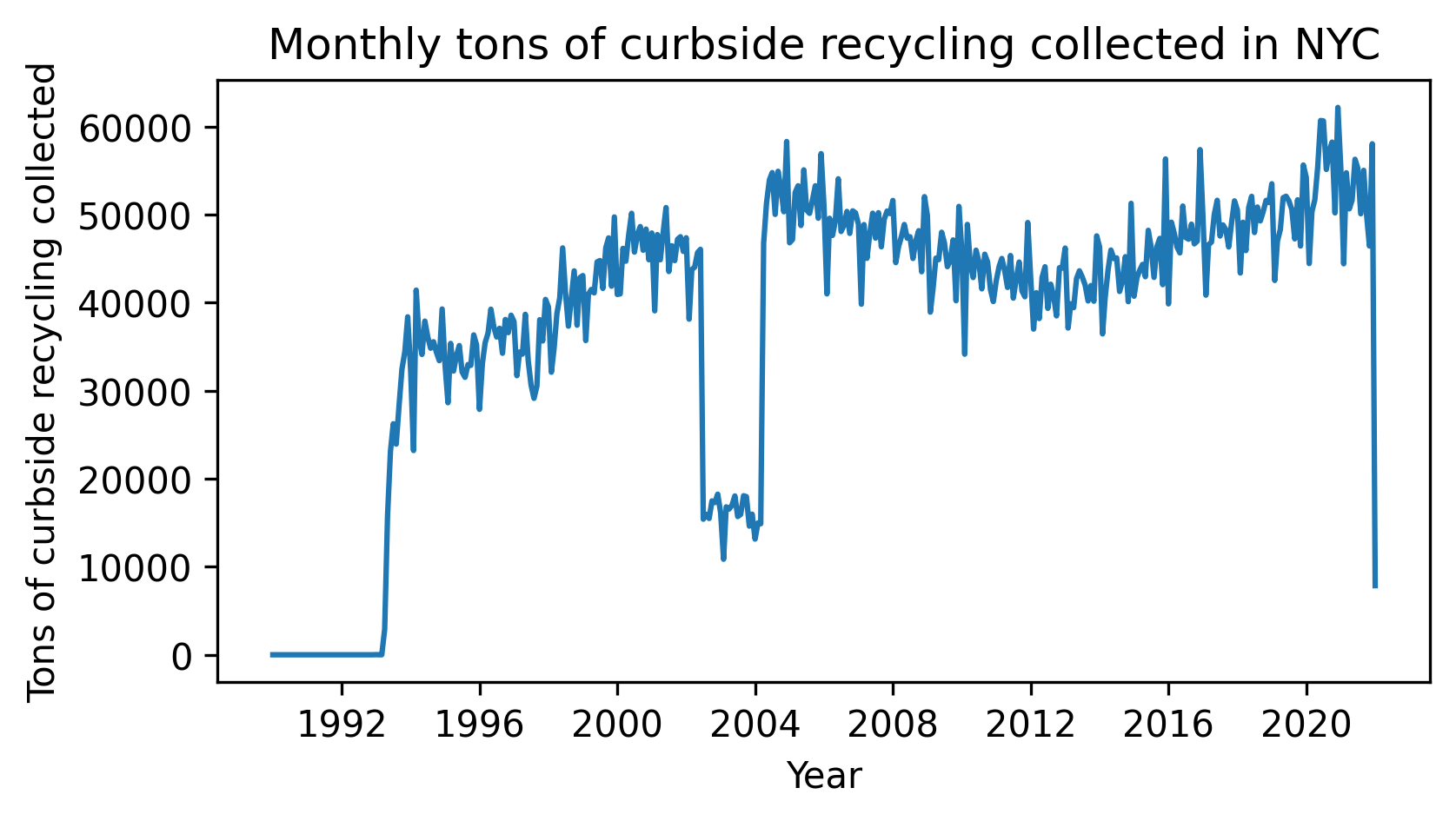}
\caption{Tons of curbside recycling collected in NYC since 1992 (full data not available before about 1993). From July 2002 to April 2004, DSNY stopped collecting some types of recycling and reduced collection frequencies. Note how, after recycling was fully resumed, the recycling quantity rebounded almost immediately.} \label{fig:recycling}
\end{figure} 
\paragraph{Assumptions}
\begin{enumerate}
    \item If composting were as accessible as recycling is now, the proportion of compost properly disposed of would approach that of recycling, without a long adjustment period. This is supported by Figure \ref{fig:recycling}, which shows how quickly recycling levels returned to normal after an extended period of decreased service. Also, NYC pilot programs which distributed bins reported high and rapidly increasing levels of composting \cite{org_pilot}.
    \item Eliminating parking spaces is not a politically tenable proposal. While it is possible, the controversy over reductions in parking \cite{barron_2019} would make any proposal involving decreasing the number of parking spaces much more difficult to pass.
    \item One cubic yard of mixed uncompacted Municipal Solid Waste weighs 250lbs or more \cite{vtw}. Thus 12yd$^3$ can hold over 1.5 tons of MSW.
    \item The zones that contain exclusively single/two family housing are R1, R2, R3-1, R3A, R3X, R4A, and R5A \cite{zones}.
    \item 12yd$^3$ of dumpster can be purchased for around $\$4155$ \cite{dumpstercost}, and residential compost bins can be purchased for $\$12$ per unit \cite{bincost}.
    \item Distribution and installation costs are no more than double the item cost. Thus compost bins are a total of no more than $\$24$ per unit and dumpsters are a total of $\$8,310$ per 12yd$^3$.
    \item Residential units produce approximately the same amount of waste. This assumption is challenged when considering e.g.~lawn waste, but generally two families will produce the same amount of waste.
\end{enumerate}
\paragraph{Solution}
We propose that 12yd$^3$ of commercially-available dumpsters be placed at fire hydrants, which are already available by law no more than 250' from every residence in the city \cite{fire_code}.
Placing dumpsters on the streets would eliminate parking spaces, but 15' to either side of a fire hydrant is already forbidden for parking \cite{dot}. 12yd$^3$ of dumpsters would only take up about 21' of curb space \cite{dumpstercost}, which would fit in the hydrant-delimited curb space with plenty of room to spare. Thus this proposal will maintain the number of parking spaces. A similar proposal to use the unused space around fire hydrants (but for bike racks) has been pursued by several partners in Philadelphia, including the Fire Department, which has similar rules regarding fire hydrants \cite{bikes}. Thus this proposal is legally feasible and compatible with the fire code. The dumpster service area will be all areas not covered by the PAYT program (i.e. all zones not exclusively single/two-family), as those areas are able to use outdoor household bins. Dumpster service will not be provided in parks, as parks do not contain high-density housing. Dumpster use will be enforced by legal changes to make it a punishable offence to dispose of waste in bags on the street if one lives in the dumpster service area. Given the DSNY's current low total enforcement costs \cite{enforcement}, we do not expect the enforcement cost to be significant.

We also propose a legislation change to apartment buildings to require them to provide easily-accessible composting locations within the buildings. This would be identical to the existing law requiring buildings to provide access to recycling bins \cite{law} and as such we do not expect the cost of enforcement (which could be done in parallel with the enforcement for recycling) to increase. Additionally, we propose distributing compost bins to single/two-family residences in NYC. Together, these changes would make composting as accessible as recycling for NYC residents.

\begin{figure}[!h]
\includegraphics[width=1\textwidth]{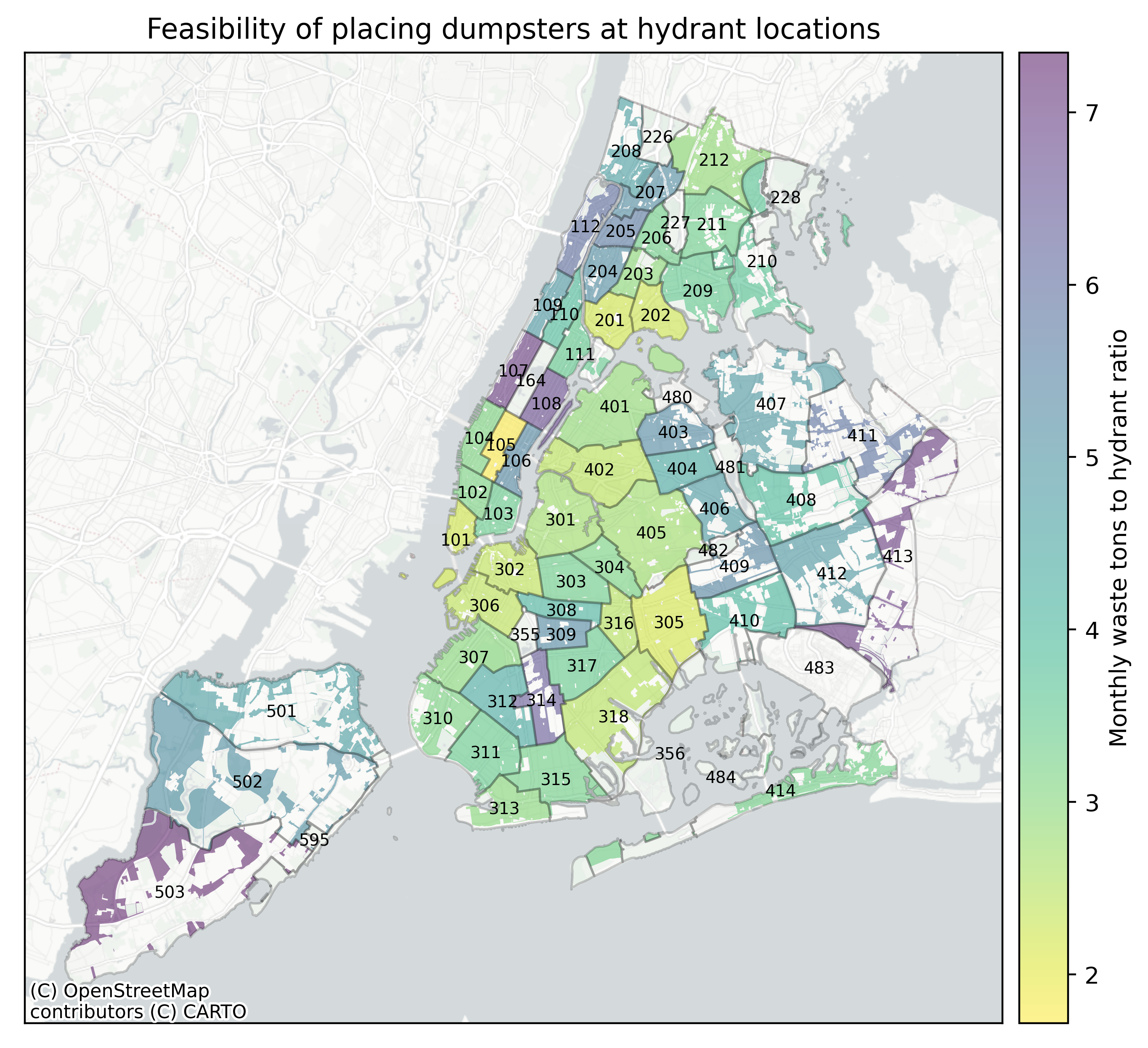}
\caption{Ratio of estimated dumpster capacity $T_i$ needed in district $i$ to number of hydrants in dumpster service areas in each community district. Community districts are labeled by 3-digit numbers, with the first digit representing the borough and the second and third digits representing the community district number.} \label{fig:ratio}
\end{figure}

\paragraph{Preliminary feasibility}
We must confirm that there are enough fire hydrants to handle the city's waste output. As NYC publishes waste tonnage data at the community district level \cite{tonnage}, we may verify for each community district that there are sufficient fire hydrants. NYC also publishes the locations of all fire hydrants in NYC \cite{hydrants}. Service area is determined by shapes specified by the NYC Department of City Planning's provided shapefiles \cite{zones-data}.
For each district $i$, we estimate the monthly waste throughput needed for the dumpsters by:
\[T_i = \frac{U_{>2,i}}{U_{i}}W_{i},\] where $T_i$ the monthly tons of waste that the dumpsters in district $i$ must accommodate, $U_{>2,i}$ is the number of lots in district $i$ with three or more residential units, $U_{i}$ is the number of residential units in district $i$, and $W_i$ the maximum tons of curbside refuse collected in one month in district $i$ since 1991. The ratio $U_{>2,i}/U_{i}$ captures the proportion of refuse generated by residences not serviced by the PAYT plan under Assumption 7. We then compute a ratio:
\[\frac{T_i}{N_i},\]
where $N_i$ is the number of hydrants in district $i$ that are in the dumpster service area. 12 yd$^3$ of dumpster (and thus one fire hydrant) can accommodate over 1.5t of MSW. Thus, the ratio $(T_i/N_i)/1.5$ represents how many times per month dumpsters would fill up in district $i$ if 12$yd^3$ of dumpsters were placed at every fire hydrant. As evidenced in Figure \ref{fig:ratio}, the highest ratio of monthly tons of waste to number of fire hydrants is around 7.35. Specifically, Community District 3 in Staten Island requires a capacity of 7.35 tons of waste per fire hydrant. So if $12yd^3$ of dumpster (with a capacity of $1.5$ tons) were placed at every fire hydrant in Staten Island's Community District 3, they would fill up 4.9 times every month. Given that curbside waste is currently collected two times a week (i.e. over 12 times per month) in the district \cite{schedule}, these dumpsters would easily handle the district's waste production.

\begin{figure}[!h]
\centering
\includegraphics[width=0.7\textwidth]{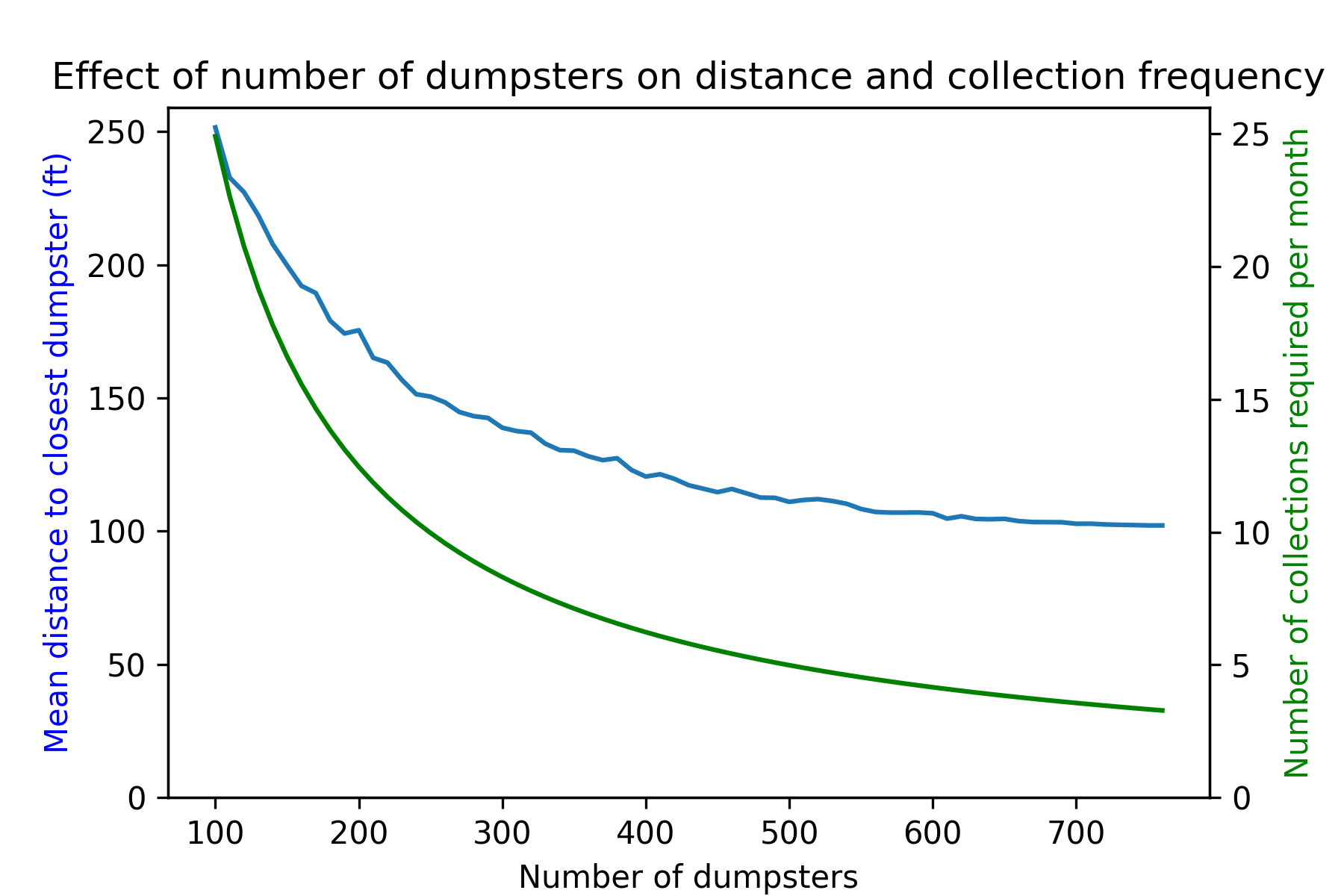}
\caption{The effect of the number of dumpsters placed in Manhattan's community district 9 on the mean distance from lots with 3 or more residential units to their closest dumpster (blue) and the required collection frequency in the district (green). Dumpster locations are determined by Algorithm \ref{algo:dumpster}, described below.} \label{fig:n_dumps}
\end{figure}

\paragraph{Choosing the number of dumpsters} As some districts have very low $T_i/N_i$ ratios, indicating that hydrant dumpster capacity far exceeds waste output, clearly a smaller number of dumpsters must be chosen in some districts. However, if the number of dumpsters is too small, the mean distance from a lot to its closest dumpster increases, as does the required collection frequency. Figure \ref{fig:n_dumps} illustrates this relationship. As much of the city currently has collection three times per week, we propose tuning the number of dumpsters in each district to accommodate three days of waste, equivalent to a required collection frequency of 10 per month.

\begin{figure}[!h]
\includegraphics[width=1\textwidth]{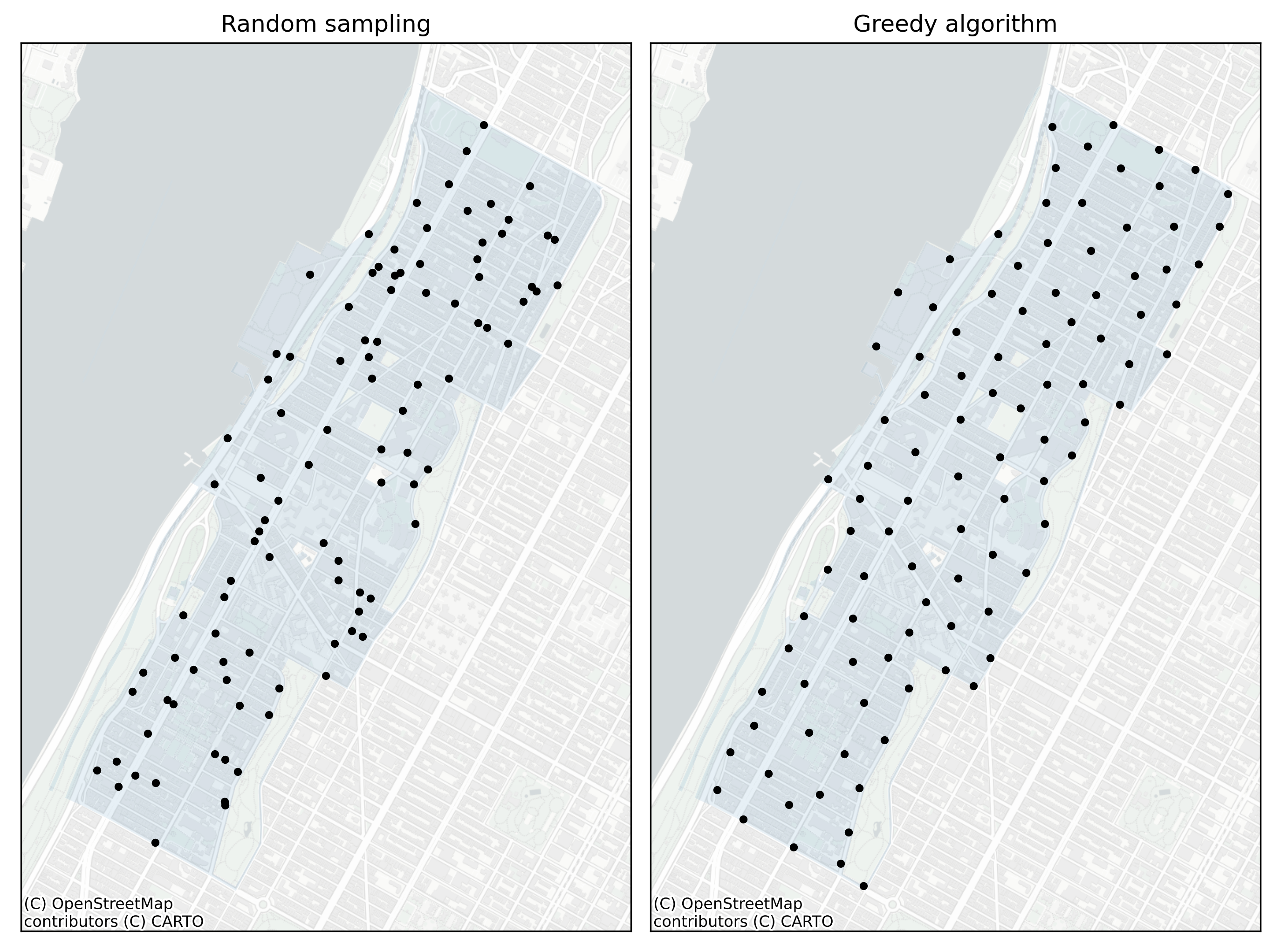}
\caption{Comparison of random sampling and the greedy algorithm as applied to Manhattan's Community District 9 (where Columbia University is located) when selecting $N=100$ dumpster locations out of 786. Note how random sampling creates areas with many dumpsters and areas with few dumpsters, which is rubbish compared to the greedy algorithm, which distributes them more evenly.} \label{fig:algo-comparison}
\end{figure}

\paragraph{Dumpster locations}
Once a number of dumpsters is chosen in a community district, we must find a way to select a subset of hydrants to place dumpsters. Ideally, this subset should be evenly-distributed throughout the area, so that no residences are extremely far from hydrants. To do this, we must compute $d_{ij}$, the distance between hydrant $i$ and hydrant $j$. We do this according to the haversine formula \cite{haversine}, which is used for computing the lengths of great circles based on the latitude and longitude of the two endpoints:
\[d_{ij}=2r\arcsin\left(\sin^2\left(\frac{\lambda_j-\lambda_i}{2}\right)+\cos(\lambda_i)\cos(\lambda_j)\sin^2\left(\frac{\mu_j-\mu_i}{2}\right)\right),\]
where $\lambda_i,\lambda_j$ are the latitudes of hydrants $i$ and $j$ and $\mu_i,\mu_j$ are the longitudes of hydrants $i$ and $j$. Next, we implement a greedy algorithm to select a well-distributed set of hydrants:

\begin{algorithm}[!htp]
\DontPrintSemicolon

Let $n$ be the number of selected hydrants, and $N$ be the target.\;
Let $H$ be the set of all hydrants.\;
Let $L$ be the set of selected hydrants.\;
Randomly select a hydrant $h_0$, set $L=\{h_0\}$, and set $n$ to 1.\;
 \While{$n$ is less than $N$}{

  For each unselected hydrant $h_i\in H-L$, compute: $D_i=\min_{j|h_j\in L}\{d_{ij}\}$, the minimum distance from $i$ to all selected hydrants.\;
  Add the hydrant with the largest $D_i$ to $L$\;
  Increment $n$ by 1.
 }
 \caption{Greedy algorithm for distributing dumpsters}
 \label{algo:dumpster}
\end{algorithm}
\noindent Effectively, this algorithm repeatedly selects the unselected hydrant that is furthest from the set of selected hydrants. Figure \ref{fig:algo-comparison} illustrates the improvement this algorithm offers over simply randomly sampling the desired number of hydrants.

\paragraph{Outcome \& Cost}
Implementing this program, we may calculate the additional tons of compost diverted under Assumption 1. The results are shown in Table \ref{table:accessibility_outcome}. Notably, Staten Island, which has less than 1/3 the population of Manhattan, is projected to divert about 59\% as much compost as Manhattan. This is likely because Staten Island produces much more lawn waste than Manhattan \cite{subsort}. Table \ref{table:accessibility_outcome} shows wide differences in the cost of the DCAP. Brooklyn, Queens, and Staten Island, which have higher proportions of lawn waste than the Bronx and Manhattan and have more single/two-family homes and can thus use cheap bins instead of expensive dumpsters, have a much lower cost per yearly tons diverted. While the cost per yearly tons diverted seems high (\$824 in Manhattan), we must keep in mind that the provision of bins and dumpsters only needs to happen once, and is not a recurring cost.

\begin{table}[!h]
\centering
\caption{Projected effects and costs of the DCAP. Current efficiency is the current proportion of curbside compost disposed of in the compost stream, calculated using DSNY's sorting data \cite{subsort}. Projected efficiency calculated under Assumption 1. Tons diverted calculated from curbside collection data for Fiscal Year 2017 \cite{tonnage}. Dumpster costs ($C_{\text{dumpsters}}$) are calculated under Assumption 6 using total dumpster numbers in each borough as calculated by the model. Bin costs ($C_{\text{bins}}$) are calculated under Assumption 6 using total numbers of units on single/two-family lots, computed from the New York Department of City Planning's MapPLUTO GIS data \cite{lots}}
\small 
\begin{tabular}[t]{r|rrrrrrrr}
\hline
Borough & Cur. eff. & Proj. eff. & $\Delta$ tons div. & $C_{\text{dumpsters}}$ & $C_{\text{bins}}$ & $C_{\text{total}}$ & $C_{\text{total}}/\Delta $ tons div. \\\hline
Bronx        &  1.2\%   &42.4\%& 66,976  & 38,874 & 2,003 & 40,877 &  0.610\\
Brooklyn     &  2.1\%   &47.4\%& 148,804 & 31,935 & 6,608 & 38,543 &  0.259\\
Manhattan    &  0.0\%    &50.0\%& 78,865  & 64,768 & 198   & 64,966 &  0.824\\
Queens       &  2.5\%   &54.4\%& 173,551 & 57,854 & 8,543 & 66,397 &  0.383\\
Staten Island&  3.1\%   &56.1\%& 46,287 & 15,515 & 3,374 & 18,888 &  0.408\\
\hline
\end{tabular}
\label{table:accessibility_outcome}
\end{table}
\subsection{Pay-As-You-Throw Program}
\paragraph{Problem} While the DCAP will aggregate trash into select drop-off location and bring composting efficiency to parity with recycling efficiency, one of the main issues plaguing NYC is the low diversion rates across all five boroughs. For the fiscal year 2021, the average diversion rate in Queens was 18.3$\%$\cite{tonnage}. Compared to other cities like San Francisco, where the rate hovers around 80$\%$\cite{frisco}, there is room for improvement in NYC.

\begin{figure}[!h]
\centering
\includegraphics[width=0.5\textwidth]{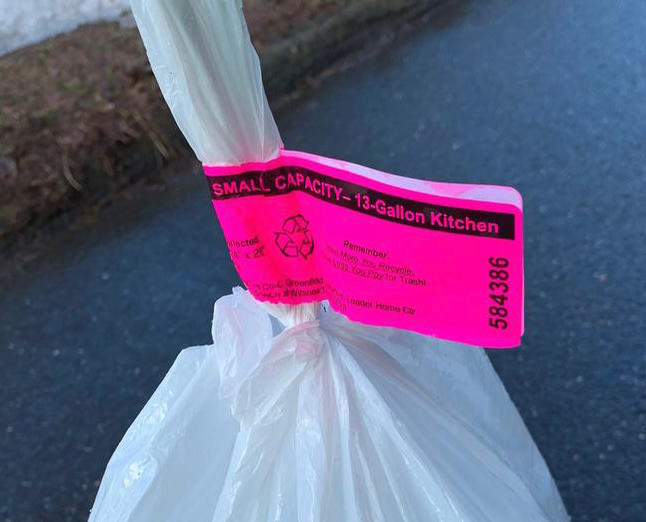}
\caption{Example sticker required for refuse bag collection. \cite{stickerpic}} \label{fig:sticker}
\end{figure}

\vspace{-3mm}

\paragraph{Solution}
Our proposed method is to enact a Pay-As-You-Throw (PAYT) program for curbside collection at single/two-family residences. This will be done by introducing stickers that residents need to purchase to have their refuse collected, as in Figure \ref{fig:sticker}. Note that this does not apply to the curbside collection of recyclables or compostable waste. We discuss this solution with reference to Queens, where 40.19\% of households (so 355,950) are located in single/two-family residences \cite{lots}. The main thing that needs to be determined is the sticker cost.

\paragraph{Model Setup} In Table \ref{table:PAYT}, we list the variables under consideration in the evaluation of a PAYT trash collection system based on increasing the price of refuse bags.

\begin{table}[!htp]
\small
\centering
\caption{PAYT model variables and constants}
\begin{tabular}[t]{cp{3.5in}rll}
\hline
Variable &\multicolumn{3}{l}{Description}&Units\\
\hline
$p$&\multicolumn{3}{l}{Price increase per 16-gallon black trash bag}&$\$$\\
$e$&\multicolumn{3}{l}{Efficiency of recycling and composting} &$\%$\\
$d_c$&\multicolumn{3}{l}{Trash disposed as compostable waste}& tons\\
$d_r$&\multicolumn{3}{l}{Trash disposed as recycling}& tons\\
$d_g$&\multicolumn{3}{l}{Trash disposed as refuse}&tons\\
$t_c$&\multicolumn{3}{l}{Total amount of compostable waste produced per year}& tons \\
$t_r$&\multicolumn{3}{l}{Total amount of recyclable waste produced per year}& tons\\
$t_g$&\multicolumn{3}{l}{Total amount of refuse produced per year} & tons \\
$C$&\multicolumn{3}{l}{Expense due to disposing trash sorted as compostable waste}&$\$$\\
$R$&\multicolumn{3}{l}{Expense due to disposing trash sorted as recyclable} &  $\$$\\
$G$&\multicolumn{3}{l}{Expense due to disposing trash sorted as refuse} &  $\$$\\
$E$&\multicolumn{3}{l}{Expense due to enforcement cost} &  $\$$\\
$T$&\multicolumn{3}{l}{Total combined expense of trash disposal} &  $\$$\\
$r$&\multicolumn{3}{l}{Revenue generated from sticker purchases} &  $\$$\\
$S_G$&\multicolumn{3}{l}{Government savings from implementation of PAYT} &  $\$$\\
$S_S$&\multicolumn{3}{l}{Societal savings from implementation of PAYT} &  $\$$\\

\hline
Constant &Description&Value&Units&Source\\
\hline
$c_c$& Cost for collecting compostable waste& 123 & $\$$/ton & \cite{allstats}\\
$c_r$ & Cost for collecting recyclable waste& 167 & $\$$/ton & \cite{allstats}\\
$c_g$ & Cost for collecting refuse & 86 & $\$$/ton & \cite{allstats}\\
$p_c$& Cost for disposing of compostable waste& 80 & $\$$/ton & \cite{allstats}\\
$p_r$ & Cost for disposing of recyclable waste& 39 & $\$$/ton & \cite{allstats}\\
$p_g$ & Cost for disposing of refuse & 30 & $\$$/ton & \cite{allstats}\\
$e_g$ & Cost for exporting refuse & 164 & $\$$/ton & \cite{allstats}\\
$L$ & Local landfill capacity for single/two family homes in Queens & 98,940& tons & \cite{lots,allstats} \\
$b_f$ & Base cost of enforcement & 16,179,000 & \$ & \cite{enforcement} \\
$b_b$ & Base cost of one trash bag & 12.35 & \$ & \cite{500household,tonnage} \\
$e_0$ & Current efficiency of composting & 2.5 & \% & \cite{subsort,tonnage}\\
$p_{c,0}$ & Current cost for collecting compostable waste & 602 & \$ & \cite{subsort,tonnage}\\
$p_{p,0}$ & Current cost of processing compostable waste & 132 & \$ & \cite{subsort,tonnage}\\
$C_0$ & Current expense due to disposing compostable waste & 351,628 & \$ & \cite{subsort,tonnage} \\
$T_0$ & Current total combined expense of trash disposal & 58,829,000 & \$ & \cite{subsort,tonnage} \\
$w$ & Weight per 16 gallon bag of trash & .01 & tons & \cite{weight}

\end{tabular}
\label{table:PAYT}
\end{table}

\noindent As the price of refuse bags increases relative to compostable waste and recycling bags, New Yorkers should be incentivized to sort their recyclable and compostable waste more to save money.  To estimate the efficiency with which New Yorkers will do so, we have used data from the PAYT system implemented in San Francisco \cite{PAYTSF} to produce a rough approximation. This system charges monthly fees for the weekly disposal of trash in primarily single-family home residential areas such that it costs $\$25.90$ per month for the weekly disposal of a single 32 gallon trash bag of refuse, while it costs $\$2.06$ per month for the weekly disposal of a single 32 gallon trash bag of either recyclables or compostable waste.  This program resulted in an $80\%$ diversion rate in 2016 \cite{frisco}. We find that  per 32 gallon bag of refuse the cost is:
\[\frac{(25.90 \, \$ / \text{month})\times (12 \, \text{months} / \text{year})}{(52 \, \text{weeks} / \text{year}) \times (1 \, \text{refuse bag} / \text{week})} \approx 5.98 \, \$ / \text{refuse bag}\]
while each 32 gallon bag of either recyclables or refuse costs:
\[\frac{(2.06 \, \$ / \text{month})\times (12 \, \text{months} / \text{year})}{(52 \, \text{weeks} / \text{year}) \times (1 \, \text{non-refuse bag} / \text{week})} \approx 0.48 \, \$ / \text{non-refuse bag}\]
Thus, a $\$5.50$ cost increase for a 32 gallon bag of refuse as compared to a 32 gallon bag of either recycling or compostable waste resulted in a $80\%$ diversion rate in San Francisco in 2016.  This can be approximated as a $\$2.75$ increase per standard $16$ gallon bag.  Based on data that roughly $90\%$ of all household waste produced in San Francisco is either compostable or recyclable \cite{frisco}, this corresponds to an effective composting and recycling efficiency of $\frac{8}{9} \approx 89\%$. As recycling efficiency is currently at $54\%$ in Queens (and, with the DCAP, we project that composting efficiency will quickly reach this rate as well) with all bags costing the same, we use a linear approximation to derive the following equation for our efficiency variable $e$ as a function of price increase $p$ per 16 gallon bag of refuse:
\[
    e(p) =
    \begin{cases}
        \frac{.89 - .54}{2.75}p + .54 & \text{if } <1  \\
        1 & \text{otherwise }  
    \end{cases}
    \]
Using this efficiency, we have the following equations relating the total amounts of each category of waste produced per year to the total amount of waste properly sorted into each category:

\[d_c(p) = e(p)\cdot t_c \, \,  \, \, \, \, \,  \, \, \, \, \,  \, \, \, d_r(p) = e(p)\cdot t_r  \,  \, \, \, \, \,  \, \, \,  \, \, \, \, \, \,  d_g(p) = t_a - (d_c(p) + d_r(p))\]

\paragraph{Calculating disposal costs} Fundamentally, disposing of trash consists of two primary costs: collection costs and processing costs.  For each class of waste, the cost per ton for both these two processes are provided in Table \ref{table:PAYT}.  For compostable waste and recyclable waste, which have the capacity to be processed within the surrounding New York City area, the total cost of disposal, a sum of the collection and processing costs, are given as follows:
\[C = (c_c+p_c)\cdot d_c(p) \, \, \, \, \, \, \, \, \, \, \, \, \, \, \, R = (c_r + p_r) \cdot d_r(p)  \]
However, for trash the situation is a bit more complicated.  New York City does not have the capacity to store all of its refuse in landfills locally, and in fact is only able to store $55\%$ of current refuse levels locally \cite{allstats}.  The rest must be exported to places such as South Carolina, adding on a substantial cost of export which must be factored into the total cost of refuse disposal.  The equation thus becomes as follows:

\[
    G(p) =
    \begin{cases}
        (p_r + p_g)\cdot d_g(p) & \text{if } d_g(p) \leq L  \\
         (p_r + p_g)L + (p_r + p_g + e_g)(d_g(p) - L) & \text{if } d_g(p) > L
    \end{cases}
    \]

\paragraph{Enforcement} Clearly, a per-bag price that is too high is not tenable. If, say, it cost \$1,000 to buy a sticker for a 16G garbage bag, there would certainly be outcomes like illegal dumping, a sticker black market, etc. Thus, in computing costs, we must take into account enforcement costs. A study in South Korea estimated that ``the legal bag price elasticity of reports on illegal dumping'' as 3.05 \cite{elasticity}. This means that a 1\% increase in the price of garbage disposal leads to a roughly 3\% increase in enforcement costs. We can thus express enforcement costs as:
\[E(p)=b_f \cdot 3.05\frac{p}{b_b},\]
where $E$ is the enforcement cost, $b_f$ is the base enforcement cost, $p$ is the price increase for a 16-gallon black trash bag, and $b_b$ is the base price of disposing of garbage. Since, in NYC, people pay for garbage collection exclusively through taxes, this base price is effectively the cost paid per bag by taxpayers for collection services.

\paragraph{Total Expense Under PAYT Program}  The total expense for trash proposal under the proposed PAYT plan is thus given by the formula:
\[T(p) = C(p) + R(p) + G(p) + E(p)\]

\paragraph{Current Total Expense} To calculate the total expense currently, which we will denote as $T_0$, we simply calculated the total current expense for composting as follows:
\[C_0 = e_0t_c(p_{c,0} + c_{c,0})\]
Unlike recycling, which occurs in excess of $50\%$ efficiency, composting is far less more widespread, and thus the cost per ton is significantly higher currently than the value assumed in the PAYT model.  This is not the case for recycling and refuse costs however, which should continue to scale accordingly per ton.  Thus, we calculate our final value $T_0$ as follows:
\[T_0 = C_0 + R(0) + G(0) + E(0)\]
which works out to roughly \$58,829,000.

\paragraph{Generated Revenue} Charging an additional fee per bag will generate added revenue for DSNY, which is calculated as follows:
\[r(p) = p\cdot \left(\frac{d_g}{w}\right)\]

\paragraph{Total Government Savings} This variable simply represents the total amount of money that the government will save based on the price per sticker as compared to the current total cost of trash disposal, taking into account revenue from the stickers themselves, and is calculated as follows:
\[S_G(p) = T_0 - (r(p) - T(p))  \]

\paragraph{Total Societal Savings} As opposed to government savings, societal savings considers that all the total revenue generated by the PAYT tax on stickers is effectively ``zero sum,''  i.e.~it is taken from NY residents, but it will be diverted back to them in some capacity.  Thus, it does not consider the added revenue stream of stickers, and is calculated as follows:
\[S_S(p) = T_0 - T(p)\]

\paragraph{Multi-Objective Optimization}
There are three conflicting possible objectives of this PAYT program: maximizing profits, minimizing societal costs, and maximizing tons of waste diverted from the refuse stream. We do not presume to know the City's preferences, thus we provide an objective function with weights to be determined by a decision-maker:
\[U(p) = \alpha S_G(p)+\beta S_S(p)+\mu (t_r+t_c),\]
where $\alpha$ is the utility assigned to one dollar of government savings, $\beta$ is the utility assigned to one dollar of societal savings, and $\mu$ is the utility assigned to diverting one ton of waste from landfills or incinerators. The parameters $\alpha, \beta, \mu$ can be obtained by asking a decision-maker the following questions:
\begin{enumerate}[label=(\Alph*)]
    \item What is the most your government is willing to spend to divert one ton of trash from landfills?
    \item What is the most your government is willing to spend to increase the wealth of your residents in single/two-family homes by \$1?
\end{enumerate}
We can then set $\alpha=1$, $\mu = A$, $\beta = B/(1-B)$. The latter equation is due to the fact that $1-B$ is the increase in societal wealth when $B$ is spent to increase the people's wealth by $1$. Once $\alpha,\beta,$ and $\mu$ are set, we use the L-BFGS-B algorithm to maximize the objective \cite{scipy}.

\section{Program Analysis}

\subsection{Dumpster and Compost Accessibility Program}
The DCAP comes out to a total cost of $\$2.30\times 10^8$. For a total diversion of $514,484$ tons of compost, this comes out to $\$446$ per yearly ton diverted. This is difficult to compare to the PAYT program or the utility function presented above, as both are in terms of yearly costs, and not a one-time payment, and as the DCAP has other benefits, including aesthetic (no more waste on the streets) and public health (possible decrease in rodent population). Nevertheless, the fact that the cost of the DCAP is a fixed, one-time cost that pales in comparison with the DSNY's $\$1.7\times 10^9$ budget, that the DCAP is legally viable, and that it requires no technological development, leads us to believe that this is a valuable proposal. To conduct a full cost-benefit analysis, more research is required into the relationship between rat populations and garbage bags left out on the street, as well as a quantification of the ``unsightliness'' of garbage bags and the economic impact of this.

\subsection{Pay-As-You-Throw Program}

\begin{figure}[!htp]
\centering
 \begin{minipage}{0.49\textwidth}
    \centering
    \includegraphics[width=1\textwidth]{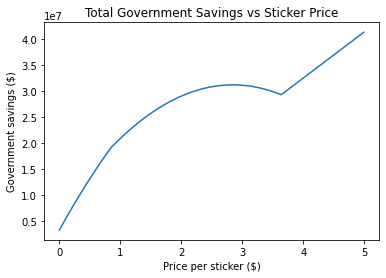}
    \caption{Total Government Savings vs. Sticker Price. A local maximum is obtained at \$2.84, where presumably diminishing trash disposal leads to a decrease in revenue generated by stickers.  Once the price reaches $\$3.63$ such that diversion efficiency is maximized, the rates of disposal methods do not change and a constant amount of stickers are bought regardless of price.  This leads government earnings to increase at a constant linear rate in line with increasing sticker prices.} \label{fig:Gov}
    \includegraphics[width=1\textwidth]{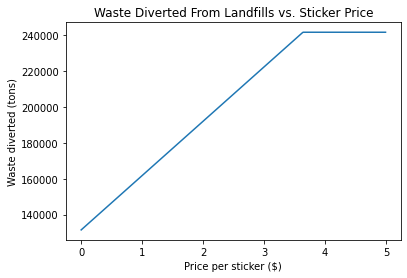}
    \caption{Waste Diverted From Landfills vs. Sticker Price. The diversion rate increases linearly with respect to price up until the price reaches $\$3.63$, at which point composting and recycling efficiency is maximized at 1 and an increase in price does not change diversion rates.\\\\}\label{fig:diversion}
 \end{minipage}\hfill
\begin{minipage}{0.49\textwidth}
        \centering
        \includegraphics[width=1\textwidth]{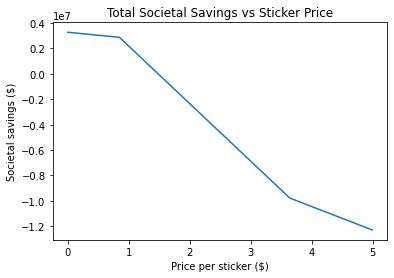}
        \caption{Total Societal Savings vs. Sticker Price.   As sticker prices increase the societal savings consistently decrease, with  approximately \$0 savings, i.e. the same cost to dispose of waste under the PAYT program as currently, when the price is  \$1.48.}\label{fig:Soc}
        \includegraphics[width=1\textwidth]{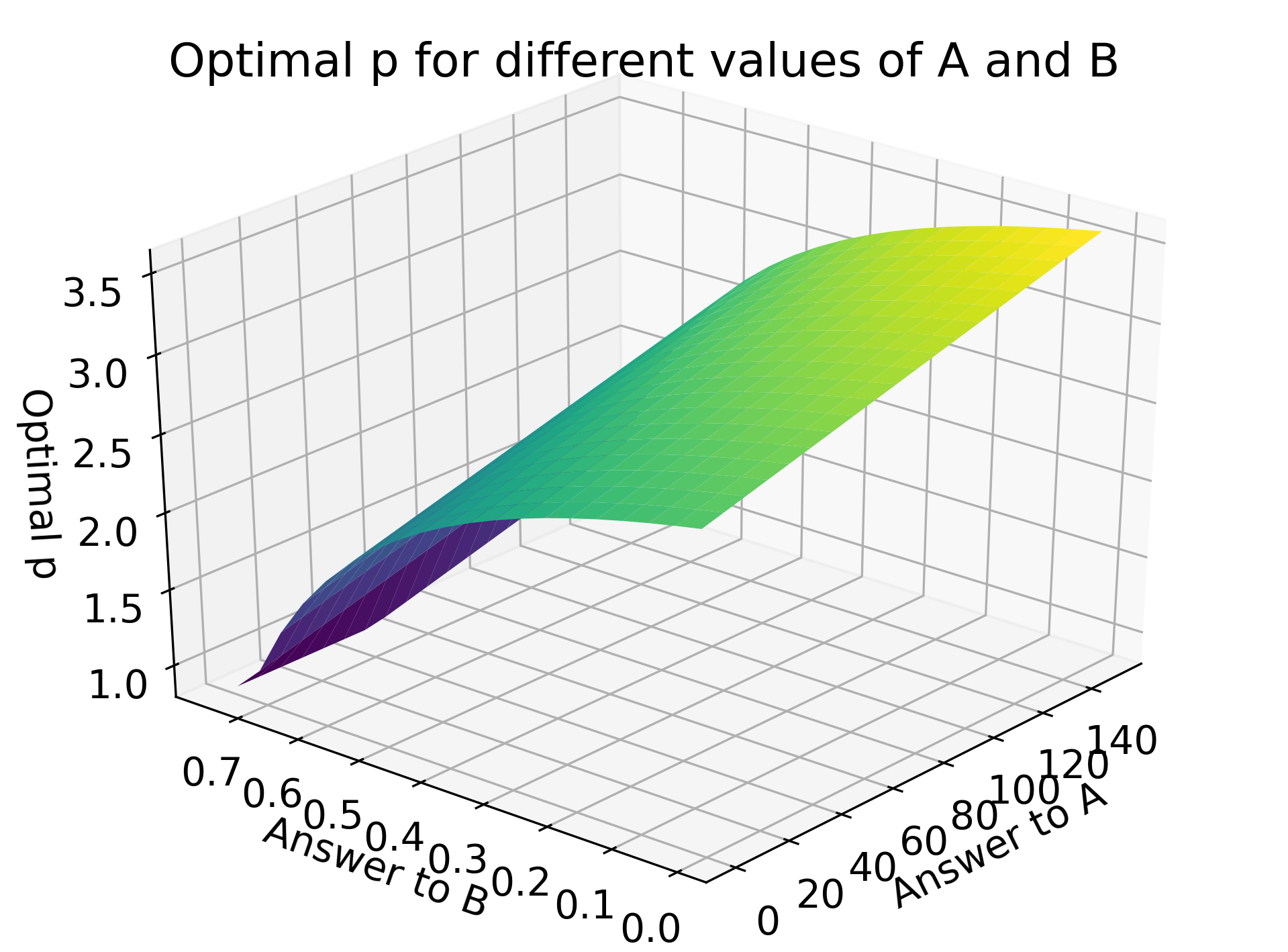}
        \caption{Plot of optimal $p$ values (i.e.~maximizing $U(p)$) for different answers to $A$ and $B$. We see that, as the answer to $A$ increases (corresponding to a higher valuation of diversion utility), $p$ increases, and that as the answer to $B$ increases (corresponding to a higher valuation of total savings in society), $p$ decreases. An ardent environmentalist who cares nothing for the wealth of society would set $p$ to extremely high values, while someone who cares nothing for the environment but thinks of residents as family would set $p$ to 0.}\label{fig:gradient}
    \end{minipage}
\end{figure}

The three factors considered to evaluate the enactment of a Pay-As-You-Throw program for all single/two-family households are its effects on diversion rates, government savings and societal savings.  Each of these factors is accounted for via sticker price, and the plots representing each of these values for sticker prices between $\$0$ and $\$5$ are shown in Figures \ref{fig:Gov}, \ref{fig:diversion} and \ref{fig:Soc}.  Additionally, for 400 pairs of answers (A,B) provided to the two proposed questions, we have determined optimal sticker pricing, and plotted these corresponding values in figure $\ref{fig:gradient}$.  

As an example of a possible policy, answering 50 dollars as A and 20 cents as B would result in an optimal sticker price of \$2.90.  This corresponds to a composting and recycling efficiency of roughly 91\% and total government savings of approximately 31 million dollars among single/two-family homes in Queens.  However, this would diminish societal savings by 64 million dollars.  

\begin{figure}[!h]
\centering
%\begin{center}
\includegraphics[width=0.8\textwidth]{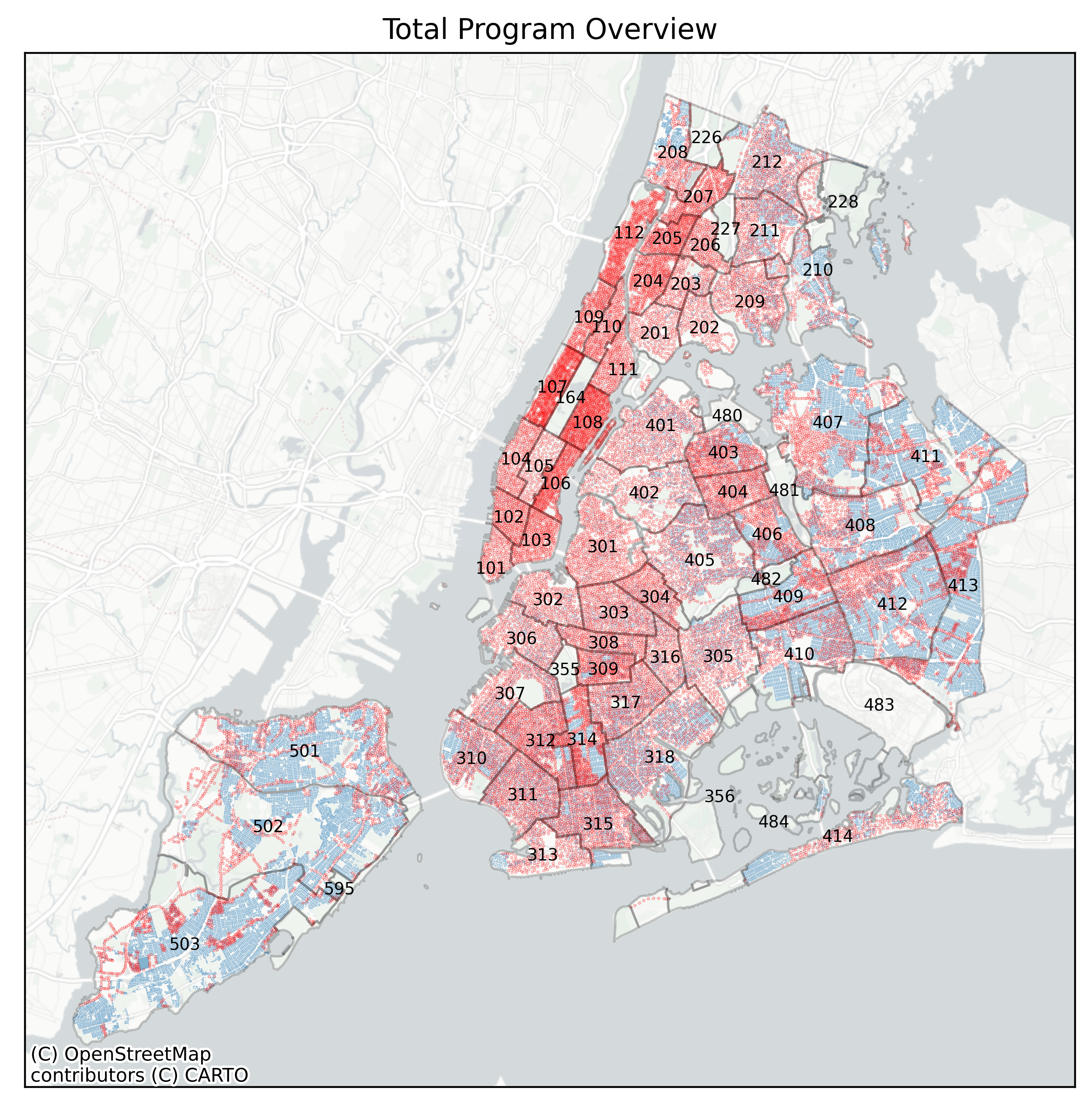}
\caption{Overview of the dumpster policy and PAYT policy. The red points are the locations of every single dumpster placed, while the blue areas are lots covered by the PAYT policy. Areas without service are areas like parks and Port of Authority of New York and New Jersey facilities.} \label{fig:policies}
%\end{center}
\end{figure}

\newpage
\section{Model Evaluation}
Having produced an analysis of the policy impacts, we may now evaluate the strengths and limitations of our models. A few strengths and limitations are common to more than one model.
\paragraph{Strengths}
\begin{itemize}
    \item Our models are simple and do not rely on very many estimated parameters.
    \item The models cleanly take into account demographic differences between different parts of NYC at the highest resolution possible (i.e. at the lot level) and without excessively discretizing the data.
\end{itemize}
\paragraph{Limitations}
\begin{itemize}
    \item Our models do not take into account the possible interaction of the DCAP and the PAYT program, as homes with PAYT service that are located very near to a dumpster may decide to use the dumpster rather than paying for stickers for curbside collection.
\end{itemize}
\subsection{Dumpster and Compost Accessibility Program}
\paragraph{Strengths}
\begin{itemize}
    \item The greedy dumpster-distribution algorithm is efficient and distributes dumpsters evenly given a certain area.
    \item The use of zoning data allows this model to position dumpsters more efficiently in each community district, as areas with single/two-family homes are already covered by the PAYT program.
    \item Assumption 1, that making composting as accessible as recycling would make composting rates quickly reach that of recycling, abstracts away the need to estimate parameters like the effectiveness and speed of cultural change, the maximum composting levels that will be reached in each borough, etc.
\end{itemize}
\paragraph{Limitations}
\begin{itemize}
    \item Certain cost data, such as the price of purchasing bins and dumpsters, may be different when the purchaser is the DSNY purchasing in bulk. This is not to mention the installation costs, which require an additional assumption to estimate, and ongoing maintenance costs, which would require even more assumptions.
    \item The zoning data is limited in that some streets, e.g.~in Staten Island's community district 3, are zoned as higher-density residential while being surrounded by single/two-family homes or parks. As the fire hydrants are located on the streets, and thus in these zones, this leads to dumpsters sometimes being placed on streets with exclusively PAYT service, or no households at all, rather than being concentrated in areas with higher-density housing (see the district labeled 503 in Figure \ref{fig:policies}). This increases the mean minimum distance to a dumpster for high-density lots. Human intervention may be necessary to select more appropriate dumpster sites in areas like this.
    \item Our cost estimates do not take into account possible collection savings due to locating waste at fewer points. Though the same volume of waste would be collected, trucks would have to stop less often (only at the dumpsters, rather than many times along a single street).
    \item Assumption 1 may not be entirely justified. Perhaps a long period of cultural change is indeed necessary for people to recycle, and the quick rebound in Figure \ref{fig:recycling} is explained by a culture persisting through a period of reduced recycling.
\end{itemize}
\subsection{Pay-As-You-Throw Program}
\paragraph{Strengths}
\begin{itemize}
    \item The objective function is computationally tractable and trivial to optimize with widely-available software.
    \item The objective function does not assume the decision-makers' preferences, and allows for any combination of marginal utilities.
    \item We provide simplified questions to help set the marginal utilities, allowing decision-makers to understand the parameters they must set. This is valuable as, in previous waste management research, the fact that some criteria were ``difficult to understand'' for decision-makers has been a problem \cite{mcda}.
\end{itemize}
\paragraph{Limitations}
\begin{itemize}
    \item We do not identify one clear ``best'' solution. Perhaps one could be arrived at with data on how much municipal governments value increasing societal wealth, as well as data on a dollar value for diverting waste from landfills (taking into account the long-term economic costs of all the environmental externalities).
    \item In the model of enforcement costs, we assume that the price elasticity of garbage bag prices on reports of illegal dumping is the same as in South Korea, which may not be justified due to differences between the two populations. Moreover, the estimate for base bag cost is flawed, as in NYC it is an ``invisible'' cost paid through taxes.
\end{itemize}

\newpage
\section{Conclusion} To address New York City's solid waste management problems, we have proposed two programs that allow us to target areas populated by both high-density apartment buildings as well as single/two-family homes.

\iffalse In a data-driven approach that combines tax lot data, and tonnage data, our model designates an appropriate number of dumpsters for each of NYC's 59 community districts, taking into account how many residents will actually be served by dumpsters and the district's waste output. A greedy algorithm then uses zoning data and data on hydrant locations to evenly distribute dumpsters over the service area.\fi
The first program is the Dumpster and Compost Accessibility Program. This program neatly solves the dearth of dumpsters in high-density areas of NYC by placing dumpsters in empty spaces surrounding fire hydrants. Though it is hard to predict the effect on rodent populations, this program proposal demonstrates that it is relatively inexpensive to resolve the problem of waste accumulating in bags on the street. By leveraging historical data on recycling, we further demonstrate that composting efficiency need not be so low: if composting were accessible, people would do it. A simple law would make composting as easy as recycling for apartment tenants, and providing bins is an inexpensive way to do the same for residents of single/two-family homes.

The second program institutes Pay-As-You-Throw policies for all residents of single/two-family homes, requiring any refuse bag to be tagged with a special sticker available for a fixed price in order to be collected.  Doing so incentivizes residents to sort out compost and recyclables from refuse to save money, increasing diversion rates.  Fundamental to this approach and its efficacy is determining an optimal sticker price.  Using data from San Francisco's enactment of a similar policy as well as data relating to the cost of garbage disposal and demographics within Queens, we developed equations to model the effects of sticker price on the efficiency of composting and recycling, savings to government spending and net societal savings.  From these equations a generalized utility function was determined, with parameters whose values are determined on a case-by-case basis using two questions relating the subjective values of each objective to decision makers.  Finally, it was demonstrated that this method of multiobjective optimization was able to produce optimal prices for several ranges of answers to the two questions posed.

To gain a full understanding of the costs and benefits of both programs, we advocate for further research into the relationship between rodent populations and bagged street trash, as well as into pricing the environmental externalities of all of the destinations of NYC's waste. To gain a better picture of optimal Pay-As-You-Throw sticker prices, we further advocate for research into the effects of sticker prices on diversion rates and illegal dumping in NYC. Since no such programs have been implemented in NYC, we suggest that several pilot programs be run in different parts of the city and with different pricing.

\renewcommand*{\bibfont}{\small}
\printbibliography

\end{document}